\documentclass[12pt]{article}
\usepackage[a4paper,width=160mm,top=25mm,bottom=25mm]{geometry}
\usepackage{amsmath}
\usepackage{esint}
\usepackage{parskip}
\usepackage{physics}
\usepackage{stix}
\usepackage{bm}
\usepackage{hyperref}%% link each equations
\usepackage{comment}
\usepackage[sort,compress,numbers]{natbib}
\usepackage{authblk} %% show author's name and location
\title{Evaluation of the applicability of Schrödinger-variant equations for classical fluid dynamics}
\author[1]{Yi-Sian Ciou}
\affil{Tainan, Taiwan} 
\numberwithin{equation}{section}% number my equations by sections
\begin{document}
\maketitle
\begin{abstract}
The artificial fluid model known as “Schrödinger flow” (SF) can represent rotational flow with dissipative effects,
 and has attracted attention despite its gap from real-world fluid behavior.
To address the structural discrepancy arising from the incomplete transition from quantum hydrodynamics to classical fluid dynamics,
 we propose a variant of the hydrodynamic Schrödinger equation (HSE) that better aligns with classical formulations.
We then revisit an alternative method for eliminating the quantum potential (referred to in this work as the ``cancellation approach'') in light of the known non-universality of the classical limit.
Beyond examining conservation laws via Lagrangian field theory,
 we identify its deeper connection to a generalized Lagrangian field theory,
 which offers a more straightforward route to constructing a Lagrangian density in terms of the components of a two-component wave function for rotational inviscid flow.
Finally, we discuss the completeness of Clebsch parameterization,
 concluding our evaluation of the applicability of the proposed SF-variant formulation.
\end{abstract}
%%%%%%%%%%%%%%%%%%%%%%%%%%%%%%%%%%%%%%%%%%%%%%%%%%%%%%%%%%%%%%%%%%%%%
\subsubsection*{Keywords}
Schrödinger equation, Lagrangian field theory, fluid dynamics, Madelung transformation, Clebsch potentials
\section{Introduction}
    From the perspective of structural similarity, the Madelung transformation\ \cite{Madelung:1927aa} serves as a bridge,
    reformulating the Schrödinger equation (SE) into a pair of equations analogous to fluid dynamics (see below).
    Such a hydrodynamic representation of quantum mechanics is termed \emph{quantum hydrodynamics}\footnote{
    In the context of QHD, ``quantum fluid'' refers to probability flow rather than material flow\ \cite[p.10]{nassar2017bohmian}.} (QHD),
    which has led to developments in various scientific fields,
    e.g., Bose-Einstein condensation\ \cite[p.8]{nassar2017bohmian} and superfluidity\ \cite{tang2023imaging};
    in particular, some researchers have redefined $\abs{\psi}^2$ as mass density,
    leading to an artificial fluid model designated as \emph{Schrödinger flow} (SF)\ \cite{meng2023quantum,chern2016smoke,chern2017fluid}.
    While it blurs the boundary between quantum and classical regimes,
    SF cannot fully represent classical fluid behavior,
    as its pressure and dissipation effect arise from spin vector,
    rather than thermodynamic processes.
    To better align SF with classical fluid dynamics,
    it is essential to analyze the underlying structural discrepancies between the two systems,
    as discussed below.
    
    At this point, we shall first introduce the Madelung transformation. The SE reads
    \begin{equation}\label{Schrödinger equation}
        {\rm i}\hbar\pdv{\psi}{t}=\widehat{H}\psi,\;
        \widehat{H} \equiv -\frac{\hbar^2}{2m}\laplacian+V,
    \end{equation}
    where $\widehat{H}$ is the Hamiltonian operator, $m$ denotes particle mass, and $V$ represents potential energy.
Consider a polar-form wave function written as $\psi(\vb{x},t)=\sqrt{A}\exp({\rm i}B/\hbar)$.
Expressing $\psi$ in terms of its amplitude and phase, the SE splits into the real and imaginary parts:
   \begin{equation}\label{Hamilton-Jacobi eq}
       \pdv{B}{t}+\frac{\abs{\grad{B}}^2}{2m} +V +Q =0,\;
       Q\equiv -\frac{\hbar^2}{2m}\frac{\laplacian(\sqrt{A})}{\sqrt{A}},
   \end{equation}
   \begin{equation}\label{continuity eq}
       \pdv{A}{t}+\divergence(A\frac{\grad{B}}{m})=0.
   \end{equation}
Eq.\ (\ref{Hamilton-Jacobi eq}) structurally resembles the Bernoulli equation for irrotational flow;
Eq.\ (\ref{continuity eq}) mimics the continuity equation with $\grad{B}/m$ being the velocity field.
The term $Q$ is recognized as a quantum potential (also called quantum potential energy; e.g.,\ \cite{rosen1986quantum}) with $A$ defined as probability density,
known as the \emph{Bohm potential}\ \cite{chern2017fluid,lopez2004nonlinear}.

Basically, by defining $A$ as mass density and requiring $[B/m]={\rm L^2 T^{-1}}$,
the fundamental postulate of SF is established; however, such a transition is incomplete.
(1)  Inheriting from quantum hydrodynamics,
the SF formulation defines $B/m$ as velocity potential (often with $m=1$ to suppress its explicit appearance, e.g.,\ \cite{meng2023quantum}).
Nonetheless,
mass of fluid enters fluid dynamics as mass density, defined continuously throughout space,
rather than mass of individual particle or molecule. In other words,
 the quantity $m$ has no utility and becomes ill-defined in this context; thus,
disengaging its involvement is suggested.
Returning to the changed meaning of $\rho$,
while $[\psi]$ changes to ${\rm M^{1/2}\rm L^{-3/2}}$, $[\widehat{H}]$ is not scaled accordingly,
which causes the dimension of functional for a wave equation no longer aligns with energy density;
in other words, the functional is not a Lagrangian density.
This transition also results in dimensional inconsistency when calculating the expectation value of an observable,
i.e., $[\int\psi^* O\psi\dd[3]{\vb{x}}]=[O]{\rm M}$, where $O$ represents an observable.
Apparently, the dimension of an operator should be changed to its original dimension per unit mass\footnote{
A scaled operator or observable is predefined rather than obtained by division by mass,
e.g., defining $w$ as enthalpy per unit mass.}.
(2) The presence of quantum potential energy diverts the SF formulation from classical fluid dynamics.
The most common device for eliminating quantum effects is to apply the classical limit, i.e., $\hbar\rightarrow 0$.
We may now pose a question: can quantum potentials be completely eliminated by reconstructing a wave equation?
In fact, it has been shown\ \cite{rosen1964relation,rosen1986quantum} that the so-called ``classical Schrödinger equation'' can be obtained by subtracting the quantum potential energy $Q$ from the Hamiltonian operator, i.e.,
\begin{equation}
    {\rm i}\hbar\pdv{\psi}{t}=\left(
        -\frac{\hbar^2}{2m}\laplacian +V -Q
    \right)\psi.
\end{equation}
We call this manipulation ``cancellation approach'' in the following.
Though this approach is less common than the classical limit,
 in some cases the classical limit is found to be non-universal;
that is, it fails when a quantum potential energy does not involve $\hbar$ (e.g.,\ \cite[Eq. (10b)]{rosen1986quantum}).
SF is no exception when represented using a two-component wave function,
thus warranting a revisit of the cancellation approach.
Most importantly, the cancellation inevitably introduces additional complexity;
therefore, it becomes essential to evaluate the applicability of SF-variant formulation.

    This paper is organized as follows. In Sec.\ \ref{Sec: Reform},
to address the issues mentioned above,
we propose the variant \emph{hydrodynamic Schrödinger equation}\footnote{We adopt the term ``hydrodynamic Schrödinger equation'' (HSE),
 originally proposed by Meng and Yang in their study on Schrödinger flow\ \cite{meng2023quantum},
as it explicitly reflects the purpose of their Schrödinger-like equation.} (HSE) and revisit the cancellation approach.
In Sec.\ \ref{Sec: L field theory},
we derive the corresponding Lagrangian density and apply Noether's theorem to examine the conservation laws.
Sec.\ \ref{Sec: Rotational inviscid flow} demonstrates the breakdown of the classical limit,
and reveals the challenge of using the cancellation approach in this scenario.
From the perspective of Lagrangian field theory, an alternative way to constructing Lagrangian density is identified.
Lastly, the completeness of Clebsch parameterization is also considered,
which completes the evaluation of the applicability of the SF-variant formulation.
Sec.\ \ref{Sec: Conclusion} concludes the paper with a summary,
briefly noting a possible extension regarding viscosity, though this direction is not pursued in the present work.

%%%%%%%%%%%%%%%%%%%%%%%%%%%%%%%

\section{The variant HSE and the cancellation approach}\label{Sec: Reform}
\subsection{The postulates of the variant HSE}\label{Sec: dimension issue and reform}
To disengage the involvement of quantity $m$,
we replace $\hbar/m$ with a dimensional factor $\phi_0$ with $[\phi_0]=[\phi]$ so that the phase is dimensionless.
An appropriate wave function in polar form is thus written as
\begin{equation}
    \psi(\vb{x},t)=\sqrt{\rho} e^{{\rm i}\phi/\phi_0},
\end{equation}
where $\rho$ represents mass density of fluid,
 and $\phi(\vb{x},t)$ denotes a velocity potential such that velocity is exactly expressed as $\grad{\phi}$ (detailed in Appendix\ \ref{Appendix: linear momentum density}).
Upon these changes, we propose the variant hydrodynamic Schrödinger equation (HSE)
 \begin{equation}\label{variant HSE}
    {\rm i}\phi_0\pdv{\psi}{t}=\widehat{\mathcal{H}}\psi,
\end{equation}
where $\widehat{\mathcal{H}}$ is designated as \emph{specific Hamiltonian operator}\footnote{The term ``specific'' is borrowed from thermodynamics, where energy per unit mass is referred to as ``specific energy''\ \cite[p.37]{graebel}.},
with a dimension of energy per unit mass. Similar to the Hamiltonian operator,
the specific Hamiltonian operator is defined as
\begin{equation}\label{specific Hamiltonian operator}
    \widehat{\mathcal{H}}\equiv-\frac{\phi_0^2}{2}\laplacian{}+ w + \Phi,
\end{equation}
where $\Phi(\vb{x},t)$ denotes potential of an external force, typically gravity, in the context of fluid dynamics;
 $w$ represents specific enthalpy of fluid (i.e., enthalpy per unit mass)\ \cite[p.4]{landau2013fluid}.
Recall the postulate that the variant HSE should disengage the involvement of the ill-defined mass $m$,
this means that $\phi_0$ is a mass-independent quantity.
Albeit $[\phi_0]=[\hbar]{\rm M}^{-1}$, $\phi_0$ is not equivalent to $\hbar/m$,
as $m$ is undefined and useless in this formulation. Lastly, concerning the value of $\phi_0$, it is a minor consideration,
as it simply approaches zero if one adopts the ``classical limit,\@'' which in this context means $\phi_0\rightarrow 0$. Alternatively,
 one may adopt the cancellation, allowing $\phi_0$ to take an arbitrary nonzero value,
  since no term involving $\phi_0$ appears in a resulting equation of motion, as will be discussed below.
%%%%%%%%%%%%%%%%%%%%%%%%%%%%%%%%%%%%%%%%%%

\subsection{The variant HSE for barotropic irrotational flow}\label{Sec: variant HSE for barotropic irrotational flow}
The Bohm potential in Eq.\ (\ref{Hamilton-Jacobi eq}) does not generally vanish due to the variable probability density.
Likewise, in this formulation, a structurally similar potential can persist due to compressibility of fluid.
 Given that such a potential is redundant,
we focus on developing the SE for compressible irrotational flow as a starting point,
and revisit the alternative approach to eliminate it beyond the classical limit.
At this juncture, we shall first briefly discuss the properties of irrotational flow.
Since irrotational flow is a consequence based on the assumption that the flow be isentropic\ \cite[p.16]{landau2013fluid},
it follows that for a compressible irrotational flow, there must be an injective relationship between mass density and pressure\ \cite[p.13]{landau2013fluid}.
In other words, such a fluid is barotropic, and thereby pressure is connected with mass density via the polytropic equation of state\ \cite{cherubini2013classical,Filippi2018VonMises,cherubini2011VonMises}
\begin{equation}\label{polytropic eq of state}
    p=K\rho^\gamma,
\end{equation}
where $p$ denotes the pressure, $K$ is a constant, and $\gamma$ is the adiabatic index\ \cite[p.114]{blundell2006thermal}.
Under these conditions, the specific enthalpy for barotropic irrotational flow takes the form
\begin{equation}\label{specific enthalpy}
    w=\int\frac{\dd{p}}{\rho}=\frac{\gamma}{\gamma-1}\frac{p}{\rho}.
\end{equation}
Since $\rho=\abs{\psi}^2$, Eq.\ (\ref{specific enthalpy}) can be rewritten as 
\begin{equation}\label{specific enthalpy resembles nonlinear term in GPE}
    w=\frac{K\gamma}{\gamma-1}\abs{\psi}^{2(\gamma-1)},
\end{equation}
which resembles the nonlinear potential in the Gross-Pitaevskii equation (GPE)\ \cite[Eq. (1)]{perez1997GPE}.

With the relations established above, we can immediately write down a basic form (i.e., not classical yet) of the SE for barotropic irrotational flow
\begin{equation}\label{variant HSE for barotropic without external potential}
    {\rm i}\phi_0\pdv{\psi}{t}=\left(
        -\frac{\phi_0^2}{2}\laplacian + \frac{\gamma}{\gamma-1}\frac{p}{\rho}
    \right)\psi,
\end{equation}
where the potential $\Phi$ is ignored for simplicity,
and this simplification is maintained throughout the following discussion.
By expressing $\psi$ in terms of its amplitude and phase, Eq.\ (\ref{variant HSE for barotropic without external potential}) splits into
\begin{equation}\label{rescaled:imaginary part}
    {\rm i}\phi_0\pdv{(\sqrt{\rho})}{t}=-\frac{{\rm i}\phi_0}{2}\left(
        2\grad{\phi}\cdot\grad(\sqrt{\rho})+\sqrt{\rho}\laplacian{\phi}
    \right),
\end{equation}
and
\begin{equation}\label{rescaled:real part}
    -\sqrt{\rho}\pdv{\phi}{t}=-\frac{\phi_0^2}{2}\left(
        \laplacian(\sqrt{\rho})-\frac{1}{\phi_0^2}\sqrt{\rho}\abs{\grad{\phi}}^2
    \right)+\frac{\gamma}{\gamma-1}\frac{p}{\rho}\sqrt{\rho}.
\end{equation}
Multiplying Eq.\ (\ref{rescaled:imaginary part}) with $2\sqrt{\rho}/{\rm i}\phi_0$ and replacing $\grad{\phi}$ with $\vb{v}$,
we obtain the continuity equation.
Dividing Eq.\ (\ref{rescaled:real part}) by $-\sqrt{\rho}$ leads to
\begin{equation}\label{Bernoulli eq involving Bohm potential}
    \pdv{\phi}{t}+\frac{1}{2}\abs{\grad{\phi}}^2+\frac{\gamma}{\gamma-1}\frac{p}{\rho}
    +\mathcal{Q}=0,\;
    \mathcal{Q}\equiv -\frac{\phi_0^2}{2}\frac{\laplacian(\sqrt{\rho})}{\sqrt{\rho}},
\end{equation}
which closely resembles the Bernoulli equation but involves a \emph{pseudo}-quantum potential $\mathcal{Q}$.
Note that we add a prefix ``pseudo-'' to distinguish the potential involving $\phi_0$ from real quantum potentials involving $\hbar$,
since this formulation is no longer quantum-mechanical in nature.
The pseudo-quantum potential has no physical significance,
conceptually similar to a “vestigial structure” inherited from quantum mechanics.
To eliminate such a redundant potential, other than the classical limit,
we can also consider canceling it by adding the negative pseudo-quantum potential to the specific Hamiltonian operator, namely,
\begin{equation}\label{variant HSE for barotropic with -Q}
    {\rm i}\phi_0\pdv{\psi}{t}=\left(
        -\frac{\phi_0^2}{2}\laplacian+\frac{\gamma}{\gamma-1}\frac{p}{\rho}
        +\frac{\phi_0^2}{2}\frac{\laplacian(\sqrt{\rho})}{\sqrt{\rho}}
    \right)\psi,
\end{equation}
which leads to the Bernoulli equation for barotropic irrotational flow\ \cite{cherubini2013classical,Filippi2018VonMises,cherubini2011VonMises}:
\begin{equation}\label{Bernoulli eq for barotropic irrotational flow}
    \pdv{\phi}{t}+\frac{1}{2}\abs{\grad{\phi}}^2+\frac{\gamma}{\gamma-1}\frac{p}{\rho}=0.
\end{equation}
Note that $\phi$ has absorbed the time-dependent constant of integration $f(t)$\ \cite{cherubini2011VonMises} such that the right-hand side of Eq.\ (\ref{Bernoulli eq for barotropic irrotational flow}) is zero;
 $f(t)$ can be recovered by writing $\phi=\tilde{\phi}-\int f(t)\dd{t}$.
Subsequently, taking gradient of Eq.\ (\ref{Bernoulli eq for barotropic irrotational flow}) gives the Euler equation
\begin{equation}
    \pdv{\vb{v}}{t}+\vb{v}\cdot\grad{\vb{v}}=-\frac{1}{\rho}\grad{p},
\end{equation}
where
\begin{equation}
    \grad(\frac{\gamma}{\gamma-1}\frac{p}{\rho})=\frac{K}{\rho}\gamma\rho^{\gamma-1}\grad{\rho}
    =\frac{1}{\rho}\grad{p}.
\end{equation}

The cancellation approach appears to provide a straightforward path to the Bernoulli equation for classical fluids\@.
This manipulation is analogous to adding a counterterm to a Lagrangian density—a crucial step in the renormalization of quantum field theory (QFT),
used to address the problem of divergences\ \cite{lancaster2014quantum}. In this context,
we regard the ``counterterm'' as a term added to a Lagrangian density to cancel the part that yields the pseudo-quantum potential.
The counterterm is expected to yield $-\mathcal{Q}$ after inserted into the Euler-Lagrange equation (ELE) for fields\@.
For the sake of rigor, we shall examine whether the cancellation leads to the conservation of classical energy and momentum,
as will be demonstrated in the next section.
%%%%%%%%%%%%%%%%%%%%%%%%%%%%%%%%%%%%%%%%%%%%%%%

\section{Lagrangian field theory}\label{Sec: L field theory}
\subsection{The effective Lagrangian density}\label{Subsec: the effective L}
The counterterm associated with $-\mathcal{Q}$ can complicate the derivation,
because it is unknown and not easily deduced.
To address this challenge, we utilize the energy-momentum tensor from Noether's theorem to identify the counterterm.
Lastly, the assumed Lagrangian density is inserted into the Euler-Lagrange equation (ELE) to verify whether it reproduces Eq.\ (\ref{variant HSE for barotropic with -Q}).
A brief review of the derivation of the ELE is provided in Appendix\ \ref{Appendix: the least action},
 whereas the derivation of energy-momentum tensor is omitted here due to the extensive algebraic manipulations required;
 interested readers are referred to the relevant literature\ \cite{cherubini2009lagrangian,greiner2000relativistic,ohanian2013gravitation}.
In this section, we do not elaborate further on these principles but instead apply them directly to streamline the demonstration below.

In analogy to the Lagrangian density for the SE\ \cite[p. 18]{greiner2000relativistic}, we assume an effective Lagrangian density
\begin{equation}\label{assumed Lagrangian}
    \mathcal{L}=
        \frac{-\phi_0^2}{2}\grad{\psi}\cdot\grad{\psi^*}
        -\frac{\phi_0}{2{\rm i}}\left(
    \psi^*\pdv{\psi}{t}-\psi\pdv{\psi^*}{t}
    \right)-\psi^* \left(\frac{K}{\gamma-1}\abs{\psi}^{2(\gamma-1)}
    \right)\psi +{C},
\end{equation}
where 
\begin{equation}\label{specific internal energy}
    \frac{K}{\gamma-1}\abs{\psi}^{2(\gamma-1)}=\frac{1}{\gamma-1}\frac{p}{\rho},
\end{equation}
is the specific internal energy (i.e., internal energy per unit mass)\ \cite{cherubini2013classical,von2012mathematical};
 $C$ is the unknown counterterm added to eliminate the pseudo-quantum potential energy density. 
To find $C$, we resort to the energy-momentum tensor
\begin{equation}\label{energy-momentum tensor}
    {T^\mu}_\nu= \pdv{\mathcal{L}}{(\pdv*{\psi_A}{x^\mu})}\pdv{\psi_A}{x^\nu}-\mathcal{L}{\delta^\mu}_\nu,\;\mu=0,\ldots,3,
\end{equation}
where we use the contravariant four-vector\footnote{Here we adopt the notation in\ \cite{cherubini2009lagrangian} where speed of light is set to 1 so that $x^0\equiv t$.} $x^\mu=\{x^0,x^i\}\equiv\{t,x,y,z\}$ with the indices $i$ running from 1 to 3\ \cite[p.3]{greiner2000relativistic},
${\delta^\mu}_\nu$ being the Kronecker symbol.
Note that $\psi_A$ represents a multiplet of scalar independent fields\ \cite[see Appendix B]{cherubini2009lagrangian}; in this context $\psi_A=(\psi,\psi^*)$.
The component ${T^0}_0$ gives the Hamiltonian density (i.e., total energy density)
\begin{equation}\label{T00}
   \begin{split}
    {T^0}_0&=\pdv{\mathcal{L}}{(\partial_t\psi)}\pdv{\psi}{t}
    +\pdv{\mathcal{L}}{(\partial_t\psi^*)}\pdv{\psi^*}{t}-\mathcal{L}
    =\frac{\phi_0^2}{2}\abs{\grad{\psi}}^2+\frac{K}{\gamma-1}\abs{\psi}^{2\gamma}-{C}\\
    &=\left(
        \frac{1}{2}\rho\abs{\vb{v}}^2 +\frac{\phi_0^2}{8}\frac{\abs{\grad{\rho}}^2}{\rho}
    \right)+\frac{p}{\gamma-1}-{C}.
   \end{split}
\end{equation}
We then deduce that ${C}={\phi_0^2}{\abs{\grad{\rho}}^2}/{8\rho}$.
To verify this assumption, we insert Eq.\ (\ref{assumed Lagrangian}) into the ELE, which, after some algebra (detailed in Appendix\ \ref{Appendix: Lagrangian density}), yields Eq.\ (\ref{variant HSE for barotropic with -Q}).
Finally, we derived the effective Lagrangian density for barotropic irrotational flow:
\begin{equation}\label{Lagrangian density for barotropic inviscid flow}
   \begin{split}
    \mathcal{L}=
    \frac{-\phi_0^2}{2}\grad{\psi}\cdot\grad{\psi^*}
    &+\frac{{\rm i}\phi_0}{2}\left(
    \psi^*\pdv{\psi}{t}-\psi\pdv{\psi^*}{t}
    \right)\\ &-\psi^* \left(\frac{K}{\gamma-1}\abs{\psi}^{2(\gamma-1)}
   -\frac{\phi_0^2}{8}\frac{\abs{\grad(\abs{\psi}^2)}^2}{\abs{\psi}^4}
    \right)\psi.
   \end{split}
\end{equation}
Notably, by expressing $\psi$ in terms of its amplitude and phase,
 Eq.\ (\ref{Lagrangian density for barotropic inviscid flow}) becomes
\begin{equation}\label{L can be written as pressure}
    \mathcal{L}=-\rho\left(
    \frac{1}{2}\abs{\vb{v}}^2 + \pdv{\phi}{t}
\right)-\frac{p}{\gamma-1}=p.
\end{equation}
In fact, this surprising result has been discovered (e.g.,\ \cite{cherubini2013classical,schakel1996effective}).
What's more, following an elaborate transformation,
Eq.\ (\ref{L can be written as pressure}) can be rewritten as\ \cite[Eq. (12)]{cherubini2013classical}
\begin{equation}\label{Lagrangian density without gravitation}
    \mathcal{L}=p=K{\left(
        \frac{1-\gamma}{K\gamma}
    \right)}^{\gamma/\gamma-1}{\left(
        \pdv{\phi}{t}+\frac{1}{2}\abs{\grad{\phi}}^2
    \right)}^{\gamma/\gamma-1}.
\end{equation}
Substituting Eq.\ (\ref{Lagrangian density without gravitation}) into the ELE results in a more complicated equation of motion,
known as the Von Mises equation\ \cite{cherubini2013classical,cherubini2011VonMises,Filippi2018VonMises}.
%%%%%%%%%%%%%%%%%%%%%%%%%%%%%%%%%%%%%%%%%%%%%%%%%%%%%%%%%%%%%

\subsection{The examination on the conservation of energy and momentum}
Given that the divergence of the energy-momentum tensor is zero, i.e.,\ \cite[Eq. (15)]{cherubini2013classical}
\begin{equation}\label{divergence of energy-momentum tensor}
    \pdv{x^\mu}({T^\mu}_\nu)=0,
\end{equation}
it provides an analytical tool for verifying whether the cancellation ensures the conservation of classical energy and linear momentum in the flow.
We begin with examining the conservation of classical energy by $\partial_\mu{T^\mu}_0=0$.
The Hamiltonian density reads (see Eq.\ (\ref{T00}))
\begin{equation}
    {T^0}_0 = \frac{1}{2}\rho\abs{\vb{v}}^2 + \frac{p}{\gamma-1}.
\end{equation}
And the components of energy flux vector ${T^i}_0$ are
\begin{equation}\label{energy flux}
    {T^i}_0=\pdv{\mathcal{L}}{(\pdv*{\psi}{x^i})}\pdv{\psi}{t} + \pdv{\mathcal{L}}{(\pdv*{\psi^*}{x^i})}\pdv{\psi^*}{t},
\end{equation}
where (see Eq.\ (\ref{ELE to barotropic-space coordinates}))
\begin{subequations}
    \begin{equation}
        \pdv{\mathcal{L}}{(\pdv*{\psi}{x^i})}
        =-\frac{\phi_0^2}{2}\pdv{\psi^*}{x^i}+\frac{\phi_0^2}{4}\frac{\psi^*}{\rho}\pdv{\rho}{x^i},
    \end{equation}
    \begin{equation}
        \pdv{\mathcal{L}}{(\pdv*{\psi^*}{x^i})}
        =-\frac{\phi_0^2}{2}\pdv{\psi}{x^i}+\frac{\phi_0^2}{4}\frac{\psi}{\rho}\pdv{\rho}{x^i}.
    \end{equation}
\end{subequations}
With these substitutes, Eq.\ (\ref{energy flux}) becomes
\begin{equation}
    \begin{split}
        {T^i}_0 &=-\frac{\phi_0^2}{2}\left(
        \pdv{\psi^*}{x^i}\pdv{\psi}{t}+\pdv{\psi}{x^i}\pdv{\psi^*}{t}
        \right)+\frac{\phi_0^2}{4}\frac{1}{\rho}\pdv{\rho}{x^i}\left(
        \psi^*\pdv{\psi}{t} + \psi\pdv{\psi^*}{t}
        \right) \\
    &=-\frac{\phi_0^2}{2}\left(
        \frac{1}{2\rho}\pdv{\rho}{x^i}\pdv{\rho}{t} + \frac{2\rho}{\phi_0^2}\pdv{\phi}{x^i}\pdv{\phi}{t}
    \right)+\frac{\phi_0^2}{4}\frac{1}{\rho}\pdv{\rho}{x^i}\pdv{\rho}{t} \\
    &=-\rho v_i\pdv{\phi}{t}.
    \end{split}
\end{equation}
With Eq.\ (\ref{Bernoulli eq for barotropic irrotational flow}), we find
\begin{equation}
    {T^i}_0=\rho{v_i}\left(
    \frac{1}{2}\abs{\vb{v}}^2+\frac{\gamma}{\gamma-1}\frac{p}{\rho}
\right).
\end{equation}
Hence, we obtain\ \cite[cf. Eq. (6.1), p.10]{landau2013fluid}
\begin{equation}
    0= \pdv{x^\mu}({T^\mu}_0)
    = \pdv{t}(\frac{1}{2}\rho\abs{\vb{v}}^2 + \frac{p}{\gamma-1})
    +\divergence[\rho\vb{v}\left(
        \frac{1}{2}\abs{\vb{v}}^2 + \frac{\gamma}{\gamma-1}\frac{p}{\rho}
    \right)].
\end{equation}
Likewise, we examine the conservation of linear momentum by $\partial_\mu {T^\mu}_i=0$.
The components of linear momentum density (see Eq.\ (\ref{T(0i)=})) read
\begin{equation}
    {T^0}_i=-\rho\pdv{\phi}{x^i},
\end{equation}
and the components of momentum flux ${T^j}_i$ are
\begin{equation}
    \begin{split}
        {T^j}_i &=\pdv{\mathcal{L}}{(\pdv*{\psi}{x^j})}\pdv{\psi}{x^i}
        + \pdv{\mathcal{L}}{(\pdv*{\psi^*}{x^j})}\pdv{\psi^*}{x^i} -\mathcal{L}{\delta^j}_i \\
        &=-\frac{\phi_0^2}{2}\left(
            \pdv{\psi^*}{x^j}\pdv{\psi}{x^i} + \pdv{\psi}{x^j}\pdv{\psi^*}{x^i}
        \right) +\frac{\phi_0^2}{4}\frac{1}{\rho}\pdv{\rho}{x^j}\left(
            \psi^*\pdv{\psi}{x^i} + \psi\pdv{\psi^*}{x^i}
        \right) -p{\delta^j}_i\\ 
        &=-\rho\pdv{\phi}{x^j}\pdv{\phi}{x^i} -p{\delta^j}_i,
    \end{split}
\end{equation}
where $\mathcal{L}$ is replaced with $p$ by Eq.\ (\ref{L can be written as pressure}).
Consequently, we obtain\ \cite[p.11]{landau2013fluid}
\begin{equation}
    0=\pdv{x^\mu}({T^\mu}_i)=\pdv{t}(-\rho v_i) + \divergence(-\rho v_i \vb{v}) -\pdv{p}{x^i}.
\end{equation}
In sum, the cancellation does ensure the conservation of classical energy and momentum.
Note that if one considers gravity, gravitational force does not appear in the momentum equation derived from Noether's theorem.
This is because gravitational potential depends explicitly on spatial coordinates,
 which breaks the invariance of $\mathcal{L}$ under spatial translations\ \cite{cherubini2009lagrangian}.
 As a result, the law of conservation of linear momentum does not hold in this scenario.
 In contrast, the conservation of energy remains valid.
%%%%%%%%%%%%%%%%%%%%%%%%%%%%%%%%%%%%%%%%%%%%%%%%%%%%%%%%%%%%%

\section{Rotational inviscid flow}\label{Sec: Rotational inviscid flow}
\subsection{Non-vanishing extra term}\label{Subsec: classical limit fails}
Return to the issue mentioned earlier that the classical limit is non-universal.
In the SF formulation, a two-component wave function is exploited to introduce vorticity\ \cite{chern2017fluid,meng2023quantum,chern2016smoke}.
Take
\begin{equation}\label{two-component wave function}
    \Psi=\mqty(\psi_1 \\ \psi_2)
    =\mqty(\sqrt{\rho(1-\alpha)}e^{{\rm i}\phi_1/\phi_0} \\ \sqrt{\rho\alpha}e^{{\rm i}\phi_2/\phi_0})
\end{equation}
with $\alpha(\vb{x},t)$ being real-valued and non-cyclic. In order to simplify the demonstration below,
 an incompressible fluid is considered, with $\rho_0$ taken as a constant mass density.
In analogy to the procedure (see Appendix\ \ref{Appendix: linear momentum density}),
the velocity is expressed as $\vb{v}=\grad{\phi_1}+\alpha\grad{\beta}$ with $ \beta\equiv \phi_2 -\phi_1$
so that it matches the Clebsch representation\ \cite{Scholle2016PhysicsLettersA,scholle2015clebsch};
 that is, $\phi_1$, $\alpha$, and $\beta$ are regarded as Clebsch potentials.
Consequently, the vorticity $\curl{\vb{v}}=\grad{\alpha}\times\grad{\beta}$ is nonzero.
The continuity equation and equation of motion are recovered from a particular combination of the real/imaginary parts of $\psi$'s.
By expressing each component in terms of its amplitude and phase, we obtain
\begin{subequations}
    \begin{equation}\label{the real part of psi_1}
        \sqrt{1-\alpha}\left(
            \pdv{\phi_1}{t} +\frac{\abs{\grad{\phi_1}}^2}{2} +\frac{p}{\rho_0}
            +\frac{\phi_0^2}{8}\frac{\abs{\grad{\alpha}}^2 +2(1-\alpha)\laplacian{\alpha}}{{(1-\alpha)}^2}
        \right)=0,
    \end{equation}
    \begin{equation}\label{the imaginary part of psi_1}
        \frac{{\rm i}\phi_0}{2\sqrt{1-\alpha}}\left[
            \left(
                \pdv{\alpha}{t} +\grad{\alpha}\cdot\grad{\phi_1}
            \right)+(\alpha-1)\laplacian{\phi_1}
        \right]=0,
    \end{equation}
\end{subequations}
for $\psi_1$, and
\begin{subequations}
    \begin{equation}\label{the real part of psi_2}
        \begin{split}
            \sqrt{\alpha}\left[
                \pdv{\phi_1}{t} +\pdv{\beta}{t} +\frac{\abs{\grad{\phi_1}}^2 +2\grad{\phi_1}\cdot\grad{\beta} +\abs{\grad{\beta}}^2}{2} +\frac{p}{\rho_0} 
                  -\frac{\phi_0^2}{8}\frac{2\alpha\laplacian{\alpha}-\abs{\grad{\alpha}}^2}{\alpha^2}
            \right]=0,
        \end{split}
    \end{equation}
    \begin{equation}\label{the imaginary part of psi_2}
        \frac{{\rm i}\phi_0}{2\sqrt{\alpha}}\left[
            \left(
                \pdv{\alpha}{t} +\grad{\alpha}\cdot\grad{\phi_1}
            \right) +(\grad{\alpha}\cdot\grad{\beta} +\alpha\laplacian{\beta}) +\alpha\laplacian{\phi_1}
        \right]=0,
    \end{equation}
\end{subequations}
for $\psi_2$.
Multiplying Eq.\ (\ref{the imaginary part of psi_1}) by $-\sqrt{1-\alpha}$ and Eq.\ (\ref{the imaginary part of psi_2}) by $\sqrt{\alpha}$,
and then adding the two resulting expressions, we recover the continuity equation
\begin{equation}
    \laplacian{\phi_1} +\grad{\alpha}\cdot\grad{\beta} +\alpha\laplacian{\beta}
    =\divergence{\vb{v}}.
\end{equation}
Likewise, multiplying Eq.\ (\ref{the real part of psi_1}) by $\sqrt{1-\alpha}$ and Eq.\ (\ref{the real part of psi_2}) by $\sqrt{\alpha}$,
and then adding the two resulting expressions, we obtain
\begin{equation}\label{Bernoulli eq with extra terms}
   \pdv{\phi_1}{t} +\alpha\pdv{\beta}{t} +\frac{\abs{\vb{v}}^2}{2} +\frac{p}{\rho_0} +\mathcal{Q}_1 +\mathcal{Q}_2
    =0,
\end{equation}
where 
\begin{equation}
    \mathcal{Q}_1 \equiv\frac{\phi_0^2}{8}\frac{\abs{\grad{\alpha}}^2}{\alpha(1-\alpha)},\;
    \mathcal{Q}_2 \equiv \frac{\alpha(1-\alpha)\abs{\grad{\beta}}^2}{2},
\end{equation}
arise from spin vector (see Appendix\ \ref{Appendix: spin-related}).
Eq.\ (\ref{Bernoulli eq with extra terms}) closely resembles the Bernoulli equation for incompressible rotational inviscid flow\ \cite[cf. Eq. (7)]{Scholle2016PhysicsLettersA}:
\begin{equation}
    \pdv{\phi}{t}+\alpha\pdv{\beta}{t} +\frac{\abs{\vb{v}}^2}{2} +\frac{p}{\rho_0}=F(\alpha,\beta,t).
\end{equation}
Note that $F$ can be eliminated by proper gauging of $\phi$, $\alpha$, and $\beta$ (detailed in\ \cite{Scholle2016PhysicsLettersA}).
As $\phi_0\rightarrow 0$, the term $\mathcal{Q}_2$ in Eq.\ (\ref{Bernoulli eq with extra terms}) still persists.
 Moreover,
$(\mathcal{Q}_1 +\mathcal{Q}_2)$ does not generally cancel out, which becomes evident when rewritten as
\begin{equation}\label{Q1 +Q2}
    \mathcal{Q}_1 +\mathcal{Q}_2 = \frac{\alpha(1-\alpha)}{2}\left[
        \abs{\grad{\beta}}^2 +{\left(
            \frac{\phi_0}{2}\frac{ \abs{\grad{\alpha}} }{ \alpha (1-\alpha) }
        \right)}^2
    \right].
\end{equation}
Since $\mathcal{Q}_2$ does not involve the dimensional factor, thereby rendering the classical limit invalid,
it suggests adopting the cancellation approach.
However, the resulting pseudo-quantum potential, $(\mathcal{Q}_1 +\mathcal{Q}_2)$,
 cannot be eliminated by directly subtracting it from $\widehat{\mathcal{H}}$,
 because it is obtained by a particular combination of the real part of expanded variant HSE for each component.
 Instead, one has to identify the relevant terms in Eq.\ (\ref{the real part of psi_1}) and Eq.\ (\ref{the real part of psi_2}) that contribute to the pseudo-quantum potentials and cancel them accordingly.
Yet, the cancellation approach in this scenario is not straightforward;
the examination via Noether's theorem becomes increasingly cumbersome. Despite this, from the perspective of Lagrangian field theory,
adding counterterms is found to be associated with a generalized theory that provides a simpler path to achieve our goal,
as demonstrated in the next section.

%%%%%%%%%%%%%%%%%%%%%%%%%%%%%%%%%%%%%%%%%%%%%%%%%%%%%%%

\subsection{Connection to a generalized theory}
Through the relationship
\begin{equation}\label{the essence of kinetic energy density}
    -\frac{\phi_0^2}{2}\abs{\grad{\psi}}^2+\frac{\phi_0^2}{8}\frac{\abs{\psi^*\grad{\psi}+\psi\grad{\psi^*}}^2}{\psi^* \psi}
    =-\frac{\rho}{2}\abs{\vb{v}}^2
    =-\frac{1}{2}\psi^*\psi\abs{
        \frac{\phi_0}{2{\rm i}}\frac{\psi^*\grad{\psi}-\psi\grad{\psi^*}}{\psi^* \psi}}^2,
\end{equation}
we recognize the connection to a generalized Lagrangian field theory by Anthony\ \cite{anthony2001hamilton}, which can describe irreversible processes in thermodynamics.
In Anthony's theory, the Lagrangian density for barotropic rotational fluid reads\footnote{Note that the dimensional factors in the phases of $\psi$ and $Z$ can be set to 1 without loss of generality,
because in Anthony's framework no extra term inherited from the quantum-like structure arises.}
\begin{equation}\label{Anthony's Lagrangian for inviscid fluid}
    \begin{split}
     \mathcal{L}= -\frac{1}{2{\rm i}}\left(
         \psi^*\pdv{\psi}{t} - \psi\pdv{\psi^*}{t}
     \right) -\frac{1}{2{\rm i}}\psi^*\psi\left(
         Z^*\pdv{Z}{t} - Z\pdv{Z^*}{t}
     \right) -\psi^* W(\psi, \psi^*)\psi \\
     -\frac{1}{2}\psi^*\psi \abs{
         \frac{1}{2{\rm i}}\frac{\psi^*\grad{\psi} - \psi\grad{\psi^*}}{\psi^*\psi} +\frac{1}{2{\rm i}}\left(
             Z^*\grad{Z} - Z\grad{Z^*}
         \right)}^2,
    \end{split}
 \end{equation}
where $\psi(\vb{x},t)=\sqrt{\rho}\exp({\rm i}\phi)$ denotes complex \emph{matter field} and $Z(\vb{x},t)=\sqrt{\Lambda}\exp({\rm i}M)$ denotes complex \emph{circulation field},
 $W$ being specific internal energy. If $Z=0$,
Eq.\ (\ref{Lagrangian density for barotropic inviscid flow}) is equivalent to Eq.\ (\ref{Anthony's Lagrangian for inviscid fluid}) with $W=K\abs{\psi}^{2(\gamma-1)}/(\gamma-1)$.
The equivalence sheds light on the fact that the cancellation is essentially rewriting classical kinetic energy density in terms of complex-valued fields.
This, in turn, suggests a simpler approach to constructing Lagrangian density for rotational flow within the SF-variant formulation,
as identifying the counterterms is less straightforward.
In light of this realization, we then transform Eq.\ (\ref{Anthony's Lagrangian for inviscid fluid}) into the form in terms of $\psi$'s:
\begin{equation}\label{Lagrangian for ideal in terms of psi's}
   \begin{split}
    \mathcal{L}=-\frac{\phi_0}{2{\rm i}}\left(
        \psi_1^*\pdv{\psi_1}{t} -\psi_1\pdv{\psi_1^*}{t} +
        \psi_2^*\pdv{\psi_2}{t} -\psi_2\pdv{\psi_2^*}{t}
    \right) -\left(
       \psi_1^*\psi_1 +\psi_2^*\psi_2
    \right)W(\psi_1,\psi_1^*,\psi_2,\psi_2^*) \\-\frac{1}{2}\left(
       \psi_1^*\psi_1 +\psi_2^*\psi_2
    \right)\abs{
        \frac{\phi_0}{2{\rm i}}\frac{(\psi_1^*\grad{\psi_1}-\psi_1\grad{\psi_1^*})+(\psi_2^*\grad{\psi_2}-\psi_2\grad{\psi_2^*})}{\psi_1^*\psi_1 +\psi_2^*\psi_2}
        }^2,
   \end{split}
\end{equation}
where $\psi_1^*\psi_1 +\psi_2^*\psi_2=\rho$.
Insertion of Eq.\ (\ref{Lagrangian for ideal in terms of psi's}) into the ELE does not directly yield a pair of Schrödinger-like wave equations for $\psi$'s,
but they can be reformulated into that form (i.e., ${\rm i}\phi_0\partial_t\psi=\widehat{\mathcal{H}}\psi+\ldots$) with the counterterms that are more complicated than $-\mathcal{Q}$.
 By expressing each component in terms of its amplitude and phase (see Eq.\ (\ref{two-component wave function})),
we obtain\ \cite[cf. Eq. (3.11)]{anthony2001hamilton}
\begin{equation}\label{Lagrangian density in terms of Clebsch potentials}
   \begin{split}
    \mathcal{L} &=-\rho\left[
        (1-\alpha)\pdv{\phi_1}{t} +\alpha\pdv{\phi_2}{t}
    \right]-\frac{1}{2}\rho\abs{(1-\alpha)\grad{\phi_1}+\alpha\grad{\phi_2}}^2 -\rho W(\rho)\\
    &=-\rho\left[\left(
        \pdv{\phi_1}{t}+\alpha\pdv{\beta}{t}
    \right) +\frac{1}{2}\abs{\grad{\phi_1}+\alpha\grad{\beta}}^2 +W(\rho)
    \right].
   \end{split}
\end{equation}
The variation of $\rho$ gives 
\begin{equation}
    \variation{\rho}:\pdv{\phi_1}{t} +\alpha\pdv{\beta}{t} +\frac{\abs{\vb{v}}^2}{2}+
    \left(
        W(\rho)+\rho\pdv{W}{\rho}
    \right)=0,
\end{equation}
where the expression within the parentheses is specific enthalpy. 
\subsection{The completeness of Clebsch parameterization}
In Anthony's theory, the amplitude/phase of each complex-valued field is independent of the others,
which allows one to introduce additional fields to satisfy specific constraints without restriction on the number of fields.
For instance,  to describe a three-dimensional non-barotropic rotational flow,
we need \emph{at least} two additional complex-valued fields in addition to the matter field $\psi$:
\begin{equation}
    Z_1=\sqrt{r}e^{{\rm i}s},\;
    Z_2=\sqrt{\lambda}e^{{\rm i}\mu},\;
\end{equation}
where $Z_1$ is associated with the conservation of entropy, i.e., $Ds/Dt=0$, where $s$ denotes specific entropy;
$Z_2$ corresponds to the Lin constraint\ \cite[p.65]{webb2018mhd}.
The velocity is then expressed as
\begin{equation}
    \vb{v}\doteq \frac{1}{2{\rm i}} \left[
        \frac{\psi^*\grad{\psi} -\psi\grad{\psi^*}}{\psi^*\psi}
        +\sum_{k=1}^{2}(Z^*_k\grad{Z_k}-Z_k\grad{Z^*_k})
    \right]=\grad{\phi} +r\grad{s} +\lambda\grad{\mu}.
\end{equation}
Note that the Lin constraint is imposed to satisfy the completeness of Clebsch parameterization (detailed in\ \cite{yoshida2009clebsch}).
Specifically, a $n$-dimensional velocity field expressed as $\grad{\phi}+\sum_{k=1}^{\nu}\alpha_k\grad{\beta_k}$ is generally complete when $\nu=n-1$.
 If the boundary values of Clebsch potentials have to be uniquely determined,
then $\nu=n$\ \cite{Webb_2016,yoshida2009clebsch}.
This implies the limited applicability of the SF-variant formulation.
A two-component wave function has only two complex-valued fields,
which gives only one pair of Clebsch potentials, i.e., $(\alpha,\beta)$.
Consequently, it can at best completely describe a two-dimensional rotational flow.
For instance, consider a two-dimensional flow under non-barotropic condition.
We assign $(r,s)\rightarrow(\alpha,\beta)$ to produce $r\grad{s}$, which implies that the conservation of entropy is imposed.
Note that $[\alpha]=1$ and $[\beta]={\rm L^2 T^{-1}}$.
To ensure dimensional consistency, we introduce a dimensional factor $r_0$ with dimension $[r_0]=[r]$,
 such that $\alpha := r/r_0$ and $\beta := r_0{s}$.
In this way,
\begin{equation}
    \alpha\grad{\beta}=\frac{r}{r_0}\grad(r_0 s)=r\grad{s}.
\end{equation}
Although the dimensions of the Clebsch potential pair in the SF-variant formulation are predetermined,
one can still assign physical quantities by this manipulation to make $\alpha\grad{\beta}$ physically meaningful.
%%%%%%%%%%%%%%%%%%%%%%%%%%%%%%%%%%%%%%%%%%%

\section{Conclusion}\label{Sec: Conclusion}
To assess the applicability of the present SF formulation to fluid dynamics,
we proposed the specific Hamiltonian operator $\widehat{\mathcal{H}}$,
which preliminary addresses the issues arising from the incomplete transition from quantum mechanics to classical fluid dynamics.
We then revisit the so-called ``cancellation approach,\@'' i.e., subtracting pseudo-quantum potential from $\widehat{\mathcal{H}}$ in this context,
since the classical limit is found to be non-universal.
By employing Noether's theorem, we validate that the cancellation leads to the conservation of classical energy and momentum.
Notably, from the perspective of Lagrangian field theory,
 the cancellation was shown to equivalent to a reformulation of classical kinetic energy in terms of complex-valued fields,
which is connected to Anthony's generalized Lagrangian field theory\ \cite{anthony2001hamilton}.
This insight provides a straightforward path to constructing a Lagrangian density,
 especially for representing rotational flow via a two-component wave function.
However, this approach inevitably introduces additional complexity,
as expressing a Lagrangian density in terms of $(\psi_1,\psi_2)$ leads to a pair of coupled, highly intricate equations,
and they need additional reformulation to recover a Schrödinger-like structure.
At the same time, the limitation of the SF-variant formulation becomes apparent when compared with Anthony's Lagrangian density,
since it employs a two-component wave function which provides only one pair of Clebsch potentials, 
thereby restricting it to completely represent rotational flows in at most two dimensions.
Moreover, even the variant HSE with a single-component wave function remains highly nonlinear (see Eq.\ (\ref{variant HSE for barotropic in terms of psi})) whenever the specific internal energy depends on the mass density of a given flow.
The variant HSE is linear and tractable only in the special case of incompressible irrotational flow, i.e.,
\begin{equation}
    {\rm i}\phi_0\pdv{\psi}{t} =\left(
        -\frac{\phi_0^2}{2}\laplacian+\frac{p}{\rho_0}
    \right)\psi.
\end{equation}

From a quantum simulation perspective, the variant HSE is not an ideal candidate for quantum algorithm development,
as linear operators are generally preferred. Several methods have been proposed for simulating classical fluids on quantum hardware,
with the lattice Boltzmann method (LBM) being a prominent candidate\ \cite{Succi_2023}.
Specifically, the LBM naturally captures both vorticity and viscosity in fluids; additionally,
the discrete-velocity Boltzmann equation\ \cite[Eq. (3.58)]{krüger2016lattice} has a structural resemblance to the Dirac equation\ \cite{SUCCI1993327},
 making the development of a quantum simulator straightforward and practical (e.g.,\ \cite{mezzacapo2015quantum}).

In contrast, within the SF-variant formulation,
one operator corresponds to one type of fluid. Regarding viscosity,
the studies\ \cite{Scholle2016PhysicsLettersA,scholle2015clebsch} may suggests a direction for extending the SF-variant formulation to incompressible viscous flow,
as they have shown that the incompressible Navier-Stokes equation can be decomposed into a Bernoulli equation with two constraint equations for a Clebsch potential pair.
While further investigation along this direction may be of mathematical interest,
its practical significance appears limited due to the considerable disadvantage discussed above.
After all, the SE was never intended for classical fluid dynamics,
and modifying it to serve this purpose yields only limited benefits.
%%%%%%%%%%%%%%%%%%%%%%%%%%%%%%%%%%%%%%%%%%%%%%%

\section*{Acknowledgment}
The author thanks Professor Alessandro Rizzo and Professor Christian Cherubini for engaging in discussions on the material presented in this paper
and deeply appreciates their generosity in sharing their time and expertise.
%%%%%%%%%%%%%%%%%%%%

\appendix

\section{The principle of least action}\label{Appendix: the least action}
It is beneficial to concisely outline the derivation of the ELE for fields.
Consider now a simply connected region $\Omega$, enclosed by a finite boundary $\partial\Omega$.
Recall that gravity was previously ignored, so $\mathcal{L}$ does not explicitly depend on spatial coordinates (i.e., $\pdv*{\mathcal{L}}{x^i}=0$)
and depends solely on $\psi_A$ and their first derivatives. The action $I$ is
\begin{equation}
    I=\int_{t_1}^{t_2}\dd{t}\int_{\Omega}\mathcal{L}(\psi_A,\partial_\mu \psi_A)\dd[3]{\vb{x}}.
\end{equation}
Varying the action with respect to $\psi_A$ gives (detailed algebraic manipulations are not shown)
\begin{equation}
   \begin{split}
    0=\variation{I}=\int_{t_1}^{t_2}\dd{t}
    \int_{\Omega} \variation{\psi_A}\left[
        \pdv{\mathcal{L}}{\psi_A} - \divergence(\pdv{\mathcal{L}}{(\grad{\psi_A})}) - \pdv{t}(\pdv{\mathcal{L}}{(\partial_t{\psi_A})})
    \right] \dd[3]{\vb{x}}\\ 
    +\int_{\Omega}\dd[3]{\vb{x}}{\left[
        \pdv{\mathcal{L}}{(\partial_t \psi_A)}\variation{\psi_A}
    \right]}_{t_1}^{t_2}
    + \int_{t_1}^{t_2}\dd{t}\int_{\Omega} \pdv{x^i}( \pdv{\mathcal{L}}{(\partial_i \psi_A)}\variation{\psi_A}) \dd[3]{\vb{x}}.
   \end{split}
\end{equation}
Following the usual manipulation\ \cite[p.481]{ohanian2013gravitation}, we impose $\variation{\psi_A}(t_1)=\variation{\psi_A}(t_2)=0$ and assume that $\variation{\psi_A}=0$ at the respective upper and lower limits of $x^i$.
 In this way, the integrals in the second line vanish. At this point,
 we shall interpret explicitly the meaning of the assumption.
Recall that in this study $\psi_A=(\psi,\psi^*)=\sqrt{\rho}\exp(\pm {\rm i}\phi/\phi_0)$, so
\begin{equation}
    \variation{\psi_A}=\pdv{\psi_A}{\rho}\variation{\rho}+\pdv{\psi_A}{\phi}\variation{\phi},
\end{equation}
which means that both $\variation{\rho}$ and $\variation{\phi}$ are assumed to vanish at the boundary.
Returning to the derivation, due to the arbitrarily varied $\variation{\psi_A}$, the expression within the brackets in the first line must vanish,
 yielding the ELE for fields:
\begin{equation}\label{ELE for fields}
    0=\pdv{\mathcal{L}}{\psi_A} - \divergence(\pdv{\mathcal{L}}{(\grad{\psi_A})}) - \pdv{t}(\pdv{\mathcal{L}}{(\partial_t{\psi_A})}).
\end{equation}
Note that even though $\psi$ and $\psi^*$ share the same quantities ($\rho$ and $\phi$), they must be varied independently\ \cite[p.18]{greiner2000relativistic}.
%%%%%%%%%%%%%%%%%%%%%%%%%%%%%%%%%%%%%%%%%%%%%

\section{The linear momentum density}\label{Appendix: linear momentum density}
For the sake of rigor, this appendix demonstrates the reform yields the correct linear momentum density.
In fact, to derive the linear momentum density, a known variant HSE under specific fluid condition is required,
analogous to how calculating the probability current needs the known SE\ \cite[p.29-30]{griffiths2019introduction}.

Recalling the continuity equation (see Eq.\ (\ref{continuity eq})),
we expand $\pdv*{\rho}{t}$ to give
\begin{equation}
    \pdv{\rho}{t}=\psi\pdv{\psi^*}{t}+\psi^*\pdv{\psi}{t}.
\end{equation}
Replacing $\pdv*{\psi}{t}$ and $\pdv*{\psi^*}{t}$ by Eq.\ (\ref{variant HSE}) and its complex conjugate,
we have
\begin{equation}
    \begin{split}
        \pdv{\rho}{t}&=\psi\left[
            \frac{-1}{{\rm i}\phi_0}\left(
                -\frac{\phi_0^2}{2}\laplacian+w+\Phi
            \right)\psi^*
        \right]+\psi^*\left[
            \frac{1}{{\rm i}\phi_0}\left(
                -\frac{\phi_0^2}{2}\laplacian+w+\Phi
            \right)\psi
        \right]\\
    &=\frac{\phi_0}{2{\rm i}}\left(
        \psi\laplacian{\psi^*}-\psi^*\laplacian{\psi}
    \right).
    \end{split}
\end{equation}
Consider
\begin{subequations}
    \begin{equation}
        \psi\laplacian{\psi^*}=
        \divergence(\psi\grad{\psi^*})-\grad{\psi^*}\cdot\grad{\psi},
    \end{equation}
    \begin{equation}
        \psi^*\laplacian{\psi}=
        \divergence(\psi^*\grad{\psi})-\grad{\psi}\cdot\grad{\psi^*}.
    \end{equation}
\end{subequations}
With these changes, the continuity equation becomes
\begin{equation}
    0=\frac{\phi_0}{2{\rm i}} \divergence(\psi\grad{\psi^*}-\psi^*\grad{\psi})
    +\divergence{\vb{J}}.
\end{equation}
Consequently,
\begin{equation}\label{redefined momentum density}
    \vb{J}=\frac{\phi_0}{2{\rm i}} \left(
    \psi^*\grad{\psi}-\psi\grad{\psi^*}
\right)=\rho\grad{\phi}.
\end{equation}
Incidentally, using the momentum-energy tensor also leads to the same result, i.e.,
\begin{equation}\label{T(0i)=}
    {T^0}_i =\pdv{\mathcal{L}}{(\partial_t\psi)}\pdv{\psi}{x^i}
        +\pdv{\mathcal{L}}{(\partial_t\psi^*)}\pdv{\psi^*}{x^i}
    =-\rho \pdv{\phi}{x^i},
\end{equation}
which correspond to the \emph{negative} components of linear momentum density.
%%%%%%%%%%%%%%%%%%%%%%%%%%%%%%%%%%%%%%%%%%%%%%%%

\section{Pseudo-quantum potentials arising from spin vector}\label{Appendix: spin-related}
The HSE derived by Meng and Yang reads\ \cite[Eq. (16), (17)]{meng2023quantum}
\begin{equation}
    \vb*{i}\hbar\pdv{\bm{\psi}}{t}=\left(
        -\frac{\hbar^2}{2}\laplacian{}+V -\frac{\hbar^2}{8\rho^2}\abs{\grad{\vb*{s}}}^2
    \right)\bm{\psi},
\end{equation}
which can also be expressed as 
\begin{equation}
    {\rm i}\hbar\pdv{t}\mqty(\psi_1 \\ \psi_2)=\left(
        -\frac{\hbar^2}{2m}\laplacian{}+V-\frac{\abs{\vb{v}}^2}{2}
        +\frac{\hbar^2}{2}\frac{\abs{\grad{\psi_1}}^2 +\abs{\grad{\psi_2}}^2}{\abs{\psi_1}^2 +\abs{\psi_2}^2}
    \right)\times\mqty(\psi_1 \\ \psi_2).
\end{equation}
Note that $\vb*{s}=\bm{\psi}^*\vb*{i}\bm{\psi}$ denotes a spin vector, and $\vb*{i}$ is a basis vector of the imaginary part of a quaternion;
$\bm{\psi}$ denotes a two-component wave function in the quaternion form, i.e.,
$\bm{\psi}=a+b\vb*{i}+(c+d\vb*{i})\vb*{j}$, where $\psi_1=a+{\rm i}b$ and $\psi_2=c+{\rm i}d$.
Comparing the Hamiltonian operators in these two forms, we can deduce that 
\begin{equation}\label{Landau-L-like potential}
    \frac{\phi_0^2}{2}\frac{\abs{\grad{\psi_1}}^2 +\abs{\grad{\psi_2}}^2}{\abs{\psi_1}^2 +\abs{\psi_2}^2}
    -\frac{1}{2}\abs{\vb{v}}^2 =\frac{\phi_0^2}{8\rho_0^2}\abs{\grad{\vb*{s}}}^2.
\end{equation}
Expressing $\psi$'s in terms of their amplitudes and phases (see Eq.\ (\ref{two-component wave function})),
we find that the left-hand side of Eq.\ (\ref{Landau-L-like potential}) is equivalent to Eq.\ (\ref{Q1 +Q2}). Hence,
\begin{equation}
    \mathcal{Q}_1 +\mathcal{Q}_2=\frac{\phi_0^2}{8\rho_0^2}\abs{\grad{\vb*{s}}}^2.
\end{equation}
%%%%%%%%%%%%%%%%%%%%%%%%%%%%%%%%%%%%%%%%%%%%%%%%%%%%%%%%%%%%%%%%%%

\section{From the Lagrangian density to the variant HSE}\label{Appendix: Lagrangian density}
Recall the assumed Lagrangian density (cf. Eq.\ (\ref{assumed Lagrangian}))
\begin{equation}
    \mathcal{L}=
        \frac{-\phi_0^2}{2}\grad{\psi}\cdot\grad{\psi^*}+\frac{{\rm i}\phi_0}{2}\left(
        \psi^*\pdv{\psi}{t}-\psi\pdv{\psi^*}{t}
    \right)-\frac{K}{\gamma-1}\abs{\psi}^{2\gamma}
        +\frac{\phi_0^2}{8}\frac{\abs{\grad{\rho}}^2}{\rho}.
\end{equation}
Note that we have replaced ${C}$ by $\phi_0^2\abs{\grad{\rho}}^2/8\rho$.
Calculate
\begin{equation}
    \pdv{\mathcal{L}}{\psi^*}=\frac{{\rm i}\phi_0}{2}\pdv{\psi}{t}-
    \left(
        \frac{K\gamma}{\gamma-1}\abs{\psi}^{2(\gamma-1)}
    \right)\psi+\pdv{\psi^*}(\frac{\phi_0^2}{8}\frac{\abs{\grad{\rho}}^2}{\rho}),
\end{equation}
where
\begin{equation}
    \begin{split}
        \pdv{\psi^*}(\frac{\phi_0^2}{8}\frac{\abs{\grad{\rho}}^2}{\rho})
        &=\frac{\phi_0^2}{8}\left(
            -\frac{\abs{\grad{\rho}}^2}{\rho^2}\psi+\frac{1}{\rho}\pdv{(\grad{\rho}\cdot\grad{\rho})}{\psi^*}
        \right)\\
        &=-\frac{\phi_0^2}{8}\frac{\abs{\grad{\rho}}^2}{\rho^2}\psi
        +\frac{\phi_0^2}{4}\frac{(\grad{\rho}\cdot\grad{\psi})}{\rho},
    \end{split}
\end{equation}
by
\begin{equation}
    \grad{\rho}=\grad(\psi\psi^*)=\psi\grad{\psi^*}+\psi^*\grad{\psi}.
\end{equation}
For $x^0$:
\begin{equation}\label{ELE to barotropic-time}
    \pdv{\mathcal{L}}{(\partial_t\psi^*)}=-\frac{{\rm i}\phi_0}{2}\psi,\;
    \pdv{t}(\pdv{\mathcal{L}}{(\partial_t\psi^*)})=-\frac{{\rm i}\phi_0}{2}\pdv{\psi}{t}.
\end{equation}
For $x^i$:
\begin{equation}\label{ELE to barotropic-space coordinates}
    \begin{split}
        \pdv{\mathcal{L}}{(\grad{\psi^*})} &=
    -\frac{\phi_0^2}{2}\grad{\psi} + \pdv{(\grad{\psi^*})}(\frac{\phi_0^2}{8}\frac{\abs{\grad{\rho}}^2}{\rho}) \\
    &=-\frac{\phi_0^2}{2}\grad{\psi} + \frac{\phi_0^2}{4}\frac{\psi}{\rho}\grad{\rho}.
    \end{split}
\end{equation}
Accordingly,
\begin{equation}
    \divergence(\pdv{\mathcal{L}}{(\grad{\psi^*})})
    =-\frac{\phi_0^2}{2}\laplacian{\psi}+\divergence(\frac{\phi_0^2}{4}\frac{\psi}{\rho}\grad{\rho}),
\end{equation}
where
\begin{equation}
    \begin{split}
        \divergence(\frac{\phi_0^2}{4}\frac{\psi}{\rho}\grad{\rho})
        &=\frac{\phi_0^2}{4}\left[
            \grad(\frac{\psi}{\rho})\cdot\grad{\rho}+\frac{\psi}{\rho}\laplacian{\rho}
        \right]\\
        &=\frac{\phi_0^2}{4}\left(
            \frac{\grad{\rho}\cdot\grad{\psi}}{\rho}-\frac{\abs{\grad{\rho}}^2}{\rho^2}\psi
        \right)+\frac{\phi_0^2}{4}\frac{\laplacian{\rho}}{\rho}\psi.
    \end{split}
\end{equation}
Finally, we obtain
\begin{equation}\label{the final step to variant HSE for barotropic}
    \begin{split}
        0&=\pdv{\mathcal{L}}{\psi^*}-\divergence(\pdv{\mathcal{L}}{(\grad{\psi^*})})
            -\pdv{t}(\pdv{\mathcal{L}}{(\partial_t\psi^*)})\\
    &={\rm i}\phi_0\pdv{\psi}{t}
        -\left(-\frac{\phi_0^2}{2}\laplacian
            +\frac{K\gamma}{\gamma-1}\abs{\psi}^{2(\gamma-1)}
        \right)\psi-\frac{\phi_0^2}{4}\left(
            \frac{\laplacian{\rho}}{\rho}-\frac{\abs{\grad{\rho}}^2}{2\rho^2}
        \right)\psi.
    \end{split}
\end{equation}
Next, since
\begin{equation}
    \frac{K\gamma}{\gamma-1}\abs{\psi}^{2(\gamma-1)}
    =\frac{\gamma}{\gamma-1}\frac{K\rho^\gamma}{\rho}=\frac{\gamma}{\gamma-1}\frac{p}{\rho},
\end{equation}
and
\begin{equation}\label{expansion of the pseudo-quantum potential}
   \frac{\phi_0^2}{4}\left(
        \frac{\laplacian{\rho}}{\rho}-\frac{\abs{\grad{\rho}}^2}{2\rho^2}
    \right)=\frac{\phi_0^2}{2}\frac{\laplacian(\sqrt{\rho})}{\sqrt{\rho}},
\end{equation}
substituting these into Eq.\ (\ref{the final step to variant HSE for barotropic}) leads to
\[0={\rm i}\phi_0\pdv{\psi}{t}
-\left(-\frac{\phi_0^2}{2}\laplacian
    +\frac{\gamma}{\gamma-1}\frac{p}{\rho}+\frac{\phi_0^2}{2}\frac{\laplacian(\sqrt{\rho})}{\sqrt{\rho}}
\right)\psi,\]
which reproduces Eq.\ (\ref{variant HSE for barotropic with -Q}). To stress its nonlinearity, we can rewrite it as
\begin{equation}\label{variant HSE for barotropic in terms of psi}
    {\rm i}\phi_0\pdv{\psi}{t}=-\frac{\phi_0^2}{2}\laplacian{\psi}+\left[
        \frac{K\gamma}{\gamma-1}\abs{\psi}^{2(\gamma-1)}+\frac{\phi_0^2}{4}\left(
            \frac{\laplacian(\abs{\psi}^2)}{\abs{\psi}^2}-\frac{\abs{\grad(\abs{\psi}^2)}^2}{2\abs{\psi}^4}
        \right)
    \right]\psi.
\end{equation}
%%%%%%%%%%%%%%%%%%%%%%%%%%%%%%%%%%%%%%%%%%%%%%%%%%%%%%%%

\bibliography{cite0511}
\end{document}